\documentclass[%
aip,%
amsmath,amssymb,floatfix,
reprint,%
jcp,%
longbibliography,
]{revtex4-1}

\usepackage{graphicx}
\usepackage{dcolumn}
\usepackage{bm}
\usepackage{amssymb}
\usepackage{multirow}
\usepackage{subfigure} 
\usepackage{xcolor}
\usepackage{mathtools}

\newcommand{\1}{\begin{equation}}
\newcommand{\2}{\end{equation}}
\newcommand{\ea}{\begin{eqnarray}} 
\newcommand{\ee}{\end{eqnarray}}
\newcommand{\4}[2]{{\frac{#1}{#2}}}

\newcommand*\chem[1]{\ensuremath{\mathrm{#1}}}

\draft 

\begin{document}


\title{Clustering-induced velocity-reversals of active colloids mixed with passive particles}



\author{Frederik Hauke}
\email[]{frederik.hauke@hhu.de}
\affiliation{Institut f\"{u}r Theoretische Physik II: Weiche Materie, Heinrich-Heine-Universit\"{a}t D\"{u}sseldorf, D-40225 D\"{u}sseldorf, Germany}
\author{Hartmut L{\"o}wen}
\email[]{hlowen@hhu.de}
\affiliation{Institut f\"{u}r Theoretische Physik II: Weiche Materie, Heinrich-Heine-Universit\"{a}t D\"{u}sseldorf, D-40225 D\"{u}sseldorf, Germany}
\author{Benno Liebchen}
\email[]{liebchen@fkp.tu-darmstadt.de}
\affiliation{Institut f\"{u}r Theoretische Physik II: Weiche Materie, Heinrich-Heine-Universit\"{a}t D\"{u}sseldorf, D-40225 D\"{u}sseldorf, Germany}
\affiliation{Institut f\"ur Festk\"orperphysik, Technische Universit\"at Darmstadt, 64289 Darmstadt, Germany.}


\date{\today}

\begin{abstract}
Recent experiments have shown that colloidal suspensions can spontaneously self-assemble into dense clusters of various internal structures, sizes and dynamical properties when doped with active Janus particles. 
Characteristically, these clusters move ballistically during their formation, but dynamically revert their velocity and 
temporarily move opposite to the self-propulsion direction of the Janus particles they contain. 
Here we explore a simple effective model of colloidal mixtures which allows 
reproducing most aspects seen in experiments, 
including the morphology and the velocity-reversal of the clusters. 
We attribute the latter to the nonreciprocal phoretic attractions of the passive particles to the active colloids' caps, 
taking place even at close contact and pushing the active particles backwards. 
When the phoretic interactions are repulsive, in turn, they cause 
dynamical aggregation of passive colloids in the chemical density minima produced by the active particles, as recently seen in experiments; in other parameter regimes they induce
travelling fronts of active particles pursued by passive ones coexisting with an active gas. 
\end{abstract}


\maketitle 

\section{Introduction}
\paragraph*{Introduction}
While many of the materials which we encounter in our environment are in equilibrium or in a glassy state, 
biological systems 
can perform a variety of functionalities, such as wound healing (self-repair) \cite{Eming2014}, mechanical adaption \cite{Weinkamer2011} 
or signaling \cite{Bieszawska2010,Beyer2018} which hinge on their nonequilibrium nature.
The latter provides biological materials with an enhanced flexibility regarding the formation of complex self-organized 
structures
as compared to passive materials whose behaviour is restricted by the detailed balance principle and the requirement to minimize free energy. 
Following biological paragons, the integration of active components into passive synthetic systems therefore offers a route towards 
fundamentally new material properties.
We are currently witnessing interesting progress in this direction, based on the doping of colloidal systems with microswimmers \cite{Paxton2004,Marchetti2013,Bechinger2016} which have been developed in the past decade \cite{Ni2013,Ni2014,Kummel2015}. 

Besides a range of theoretical studies focusing on motility-induced phase separation in active-passive mixtures
\cite{Stenhammar2015,Wittkowski2017,Wysocki2016,Dolai2018} occurring at moderate densities, recent experiments 
\cite{Singh,Wang2018,Wang2019,ibele2010,WangExclusion}
have observed the formation of various self-assembled structures even at very low active particle densities. Since these structures resemble those of ordinary molecules, the emerging clusters are called heteronuclear active colloidal molecules \cite{LoewenEPL2018,Ebbens2016,Mallory2018,Soto2014PRL,Soto2015PRE,Niu2017,Schmidt2019,Cheng2014,Wang2013,Wang2015,Gao2017,Ni2017,Kapral2007,Niu2017PRL}.
Ref.~\cite{Singh} in particular has explored passive colloids doped with light-activated active Janus colloids (active packing fraction $\phi_a=0.4\%$), 
reporting the aggregation of passive colloids around active 'seeds', even at densities which are too low to allow the 
active colloids to aggregate in the absence of the passive ones. This has recently been modeled in Ref. \cite{Stürmer2019} describing a similar aggregation dynamics 
as seen in experiments.
However, in experiments \cite{Singh}, at early stages, the emerging clusters move ballistically, like the active colloids they contain, but interestingly, 
they temporarily move in the opposite direction, representing a velocity-reversal (not described in Ref. \cite{Stürmer2019}). 

In the present work, we explore a simple model for chemically-interacting active-passive colloidal mixtures reproducing most aspects seen in the experiments \cite{Singh}. They range from the initial aggregation of passive particles around the active seeds
at active-particle densities which are too low to allow them aggregating alone, 
to the clustering-induced velocity-reversals and the 
overall morphology of the resulting patterns at intermediate and late times. 
The model hinges on the well-known fact that both active and passive particles move up or down the chemical gradients produced by the active particles 
\cite{Anderson1989,Saha2014,Pohl2014,Liebchen2015,Liebchen2017,Huang2017,Stark2018,Liebchen2018,Kapral2018,Canalejo_Golestanian,Grauer} representing chemical (or 'phoretic') interactions. The model accounts for both, the nonreciprocal and the anisotropic character of these interactions, both being necessary to describe the clustering-induced velocity-reversals seen in experiments. 
Phoretic interactions may therefore be sufficient to 
describe the key aspects of the active-passive colloidal mixtures explored in \cite{Singh}, although hydrodynamic interactions, which we neglect here, 
may generally be also important for some (other) active colloids and active-passive mixtures \cite{Liebchen2019} 
(see also the interesting very recent flow field measurements in ref.~\cite{Campbell2019}).

Other recent experiments \cite{Wang2018,Wang2019}
have revealed that under appropriate conditions active colloids may also (phoretically) repel 
passive colloids (e.g. \chem{TiO_2} Janus colloids under UV-light illumination).
For such mixtures, our model predicts a rich set of possible patterns involving 
states where passive colloids dynamically aggregate in the minima of the phoretic field produced by the active particles as well 
as rigid core-shell structures and ballistically traveling fronts of active and passive particles moving through a background gas of active colloids. Unlike other patterns predicted in this work, the traveling fronts have not yet been observed experimentally and might therefore inspire corresponding new work. 

\begin{figure}
  \centering
  \includegraphics[width=0.50\textwidth]{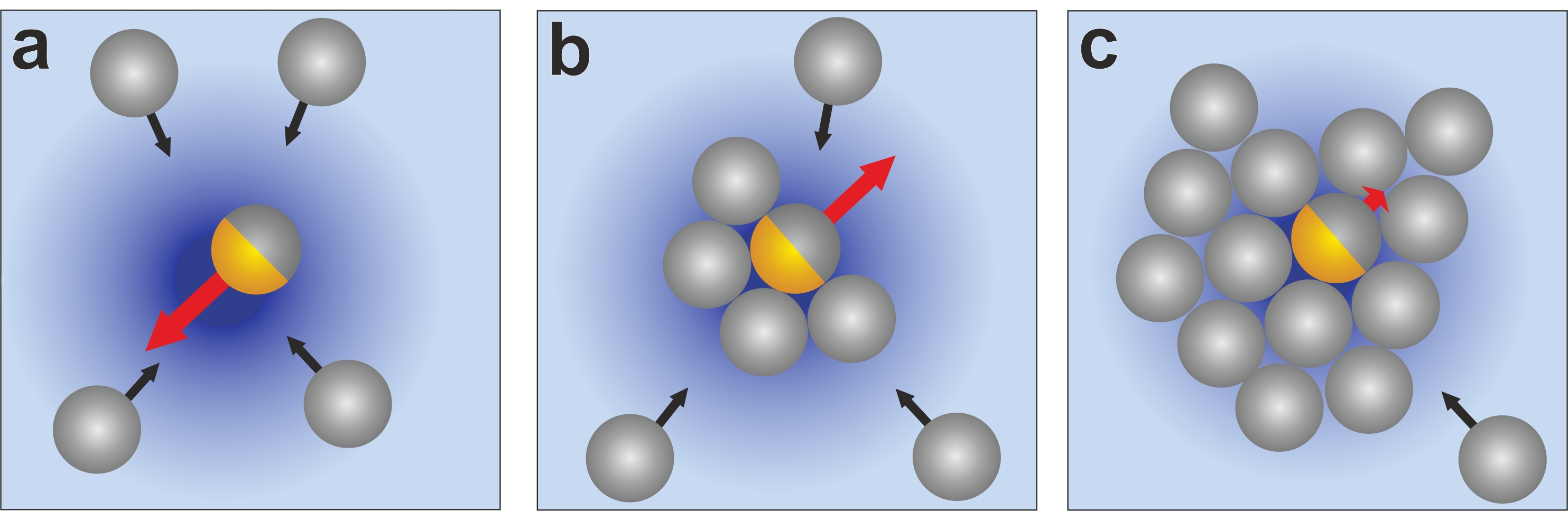}
  \caption{Schematic illustration of cluster formation and clustering-induced velocity-reversals.
  a) Initially the active colloids self-propel with their catalytic cap (golden) ahead, i.e. up their self-produced chemical gradients due to (diffusio)phoresis. 
  Passive colloids (grey) phoretically move up the same gradients towards the cap of the active particles. 
(b) When being attaching to the cap of an active colloid, the passive colloids continue moving up the chemical gradient and collectively push 
the active particle against its self-propulsion direction, resulting in a reversed cluster-velocity. 
c) As the clusters grow and the active particle is surrounded by at least one complete shell of passive colloids it becomes very slow. }
\label{Cartoon}
\end{figure}

\section{Model}\label{model} 
We describe a mixture of $N_p$ passive colloids and $N-N_p$ self-propelling active colloids
of identical diameter $d$ and with translational $D_t$ and rotational $D_r$ diffusion coefficient. Each active colloid catalyzes a certain chemical 
reaction on a part of its surface which leads to self-propulsion by (diffusio)phoresis and causes 
chemical cross-interactions due to the (diffusio)phoretic response of all other colloids to the resulting chemical field. 
To model the many-particle dynamics of the considered mixture, we formulate a simple effective model based on a similar framework as underlying previous works, such as 
\cite{Saha2014,Pohl2014,Liebchen2017,Liebchen2019} (see Appendix for details), 
but in a somewhat different form accounting for aspects which are important to describe  
experiments \cite{Singh} (e.g.~the anisotropic chemical production of the active colloids) \footnote{The recently developed AAA model \cite{Liebchen2019} could be used as an alternative to the present model, after extending it to account for the anisotropy in the chemical production as necessary to model the velocity-reversals; however it does not admit to explore delay-effects due to the noninstantaneous
chemical relaxation, which can be important for the case of repulsive phoretic interactions \cite{Liebchen2015,Liebchen2017} and which we study here for active-passive mixtures.}: 

The $N_p$ isotropic passive colloids do not self-propel but respond to the gradient of the chemical field $c({\bf r},t)$ 
produced by the active colloids. We model their dynamics using Langevin equations (in dimensionless form) 
\begin{align}  
\dot{{\bf r}}_i &=& \mathcal B_p \nabla c\left.\right|_{{\bf r}={\bf r}_i}  - \nabla_{{\bf r}_i} \mathcal{U} + \sqrt{2\mathcal{D}_t} {\boldsymbol {\xi}}_i; \quad i=1,..,N_p \label{eq:p} 
\end{align}
Here we have used the inverse rotational diffusion coefficient 
$1/D_r$ and the particle diameter $d$ as our time and space units respectively and have introduced
$\mathcal D_t = \frac{D_t}{D_r d^2}$ and the reduced coupling strength 
$\mathcal B_{p} = \frac{\beta_{p}}{D_r d^{D+2}}$ to the chemical field $c({\bf r},t)$, which is produced by the active colloids. $\mathcal{\beta}_{p}$ is the bare coupling strength and $D$ the dimensionality of the system. ${\boldsymbol{\xi}}_i(t)$ describes Gaussian white noise of zero mean and unit variance.
$\tilde U$ represents the steric interaction potential in dimensionless form and is modeled here using Weeks-Chandler-Anderson-interactions 
$\mathcal{U} = \sum_{i<j} \tilde u(r_{ij})$ \cite{WCA} where $r_{ij}=|{\bf r}_i-{\bf r}_j|$ and
\begin{equation}
\tilde u(r_{ij}) =  \4{\epsilon}{\gamma d^2 D_r} 
\left\{4 \left[\left(\frac{\sigma}{r_{ij}}\right)^{12}-\left(\frac{\sigma}{r_{ij}}\right)^6\right] + 1 \right \}
\end{equation}
if $r_{ij}/\sigma \leq 2^{1/6}$ and $\tilde u(r_{ij})=0$ else.
Here $\sigma=d/2^{1/6}$, $\epsilon$ is an energy, defining the interaction strength, and $\gamma$ is the Stokes drag coefficient.

\begin{figure}
  \centering
  \includegraphics[width=0.40\textwidth]{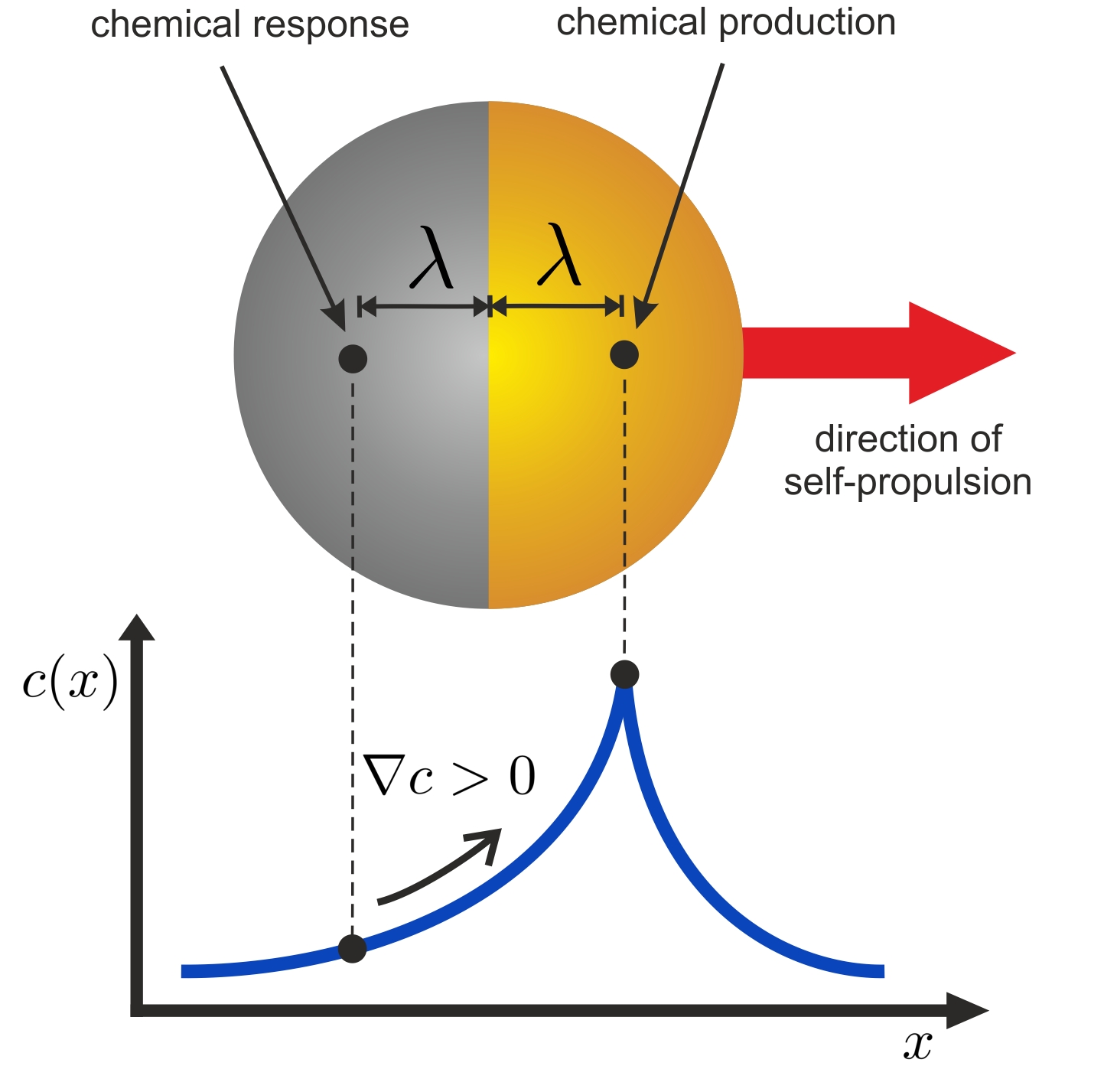}
  \caption{Schematic illustration of an active colloids as described in the present effective model: The chemical production 
  on the cap is replaced by a single point source at the 'center' of the cap which has a distance $\lambda=\frac{2}{3\pi}d$ from the geometrical center of the colloids. 
  In the model each (anisotropic) active colloid sees the chemical gradient at a single point slightly displaced from 
  its center, whereas isotropic passive colloids see the gradient in their center. 
  Accordingly, both anisotropic phoretic interactions and self-propulsion emerge from the same chemical field.}
\label{AnisoProd}
\end{figure}

To model the dynamics of the active colloids, we use the following effective Langevin equations 
\begin{align}  
\dot{{\bf r}}_i &=& \mathcal B_a \nabla c \left.\right|_{{\bf r}={\bf r}'_i} - \nabla_{{\bf r}_i} \mathcal{U} + \sqrt{2\mathcal{D}_t}
{\boldsymbol {\xi}}_i \label{eq:I1} \\
\dot{\theta}_i &=& \sqrt{2} \,\eta_i  ;\quad \quad i=N_p+1,..,N, \label{eq:IIb}
\end{align}
Here we have introduced the reduced coupling coefficient of the active colloids to the chemical field, $\mathcal B_{a} = \frac{\beta_{a}}{D_r d^{D+2}}$, where 
$\beta_{a}$ is the bare coupling coefficient, and ${\bf r}'_i={\bf r}_i-\lambda {\bf p}_i$ represents a small shift from the midpoint of colloid $i$ leading to self-propulsion (see Fig.~\ref{AnisoProd}). 
That is, we assume that the isotropic passive colloids effectively see the chemical field at their midpoint and that the anisotropic active Janus colloids 
see the field at some position shifted to their midpoint. 
This simple picture does of course not describe the actual microscopic details of self-propulsion (therefore we call the model 'effective'), but leads to 
the correct phoretic far-field interactions and additionally
accounts qualitatively for the asymmetry of the chemical interactions. The latter  
is necessary to describe the velocity-reversals seen in experiments
when active and passive colloids come together. 
Note also that Eq.~(\ref{eq:IIb}) neglects phoretic alignment effects, 
because the experiments in Ref.~\cite{Singh} have been performed at very low active-particle concentrations and are dominated by active-passive phoretic interactions for which alignment effects do not occur, not by active-active ones.

The effective chemical field follows a chemical diffusion equation supplemented by a point source for each active colloid, placed at the geometric center of the cap (projected to 2D), and a sink term accounting for a decay of the chemical field, probably occurring due to 
bulk reactions which might be triggered by the UV-illumination of the system in experiment \cite{Singh,Wang2018,Wang2019,ibele2010,WangExclusion}. 
In dimensionless form, the equation of motion for the chemical dynamics reads:
\begin{equation} \label{eq:I3}
\dot{c}({\bf r}, t) = \mathcal{D}_c \nabla^2 c + K_0\sum_{i \in \{active\}} \delta({\bf r}-[{\bf r}_i(t)+\lambda {\bf p}_i]) - K_d c
\end{equation}
Here we have introduced the reduced chemical diffusion coefficient 
$\mathcal D_c = \frac{D_c}{D_r d^2}$, the reduced production rate $K_0=\frac{k_0}{D_r}$ and the reduced decay rate $K_d=\frac{k_d}{D_r}$ and do not assume that the chemical field follows the position of the colloids instantaneously, which is particular relevant when the phoretic interactions are repulsive and cause delay effects, as we will see. 
Note that the present model neglects hydrodynamic interactions which may 
generally be or not be important depending on the specific active colloids under consideration  
\cite{Liebchen2019}, but are not needed to 
reproduce the phenomenomena occuring in experiment \cite{Singh} within our model.


\section{Chemical attractions}\label{chemicalattractions}
\begin{figure}
  \centering
  \includegraphics[width=0.45\textwidth]{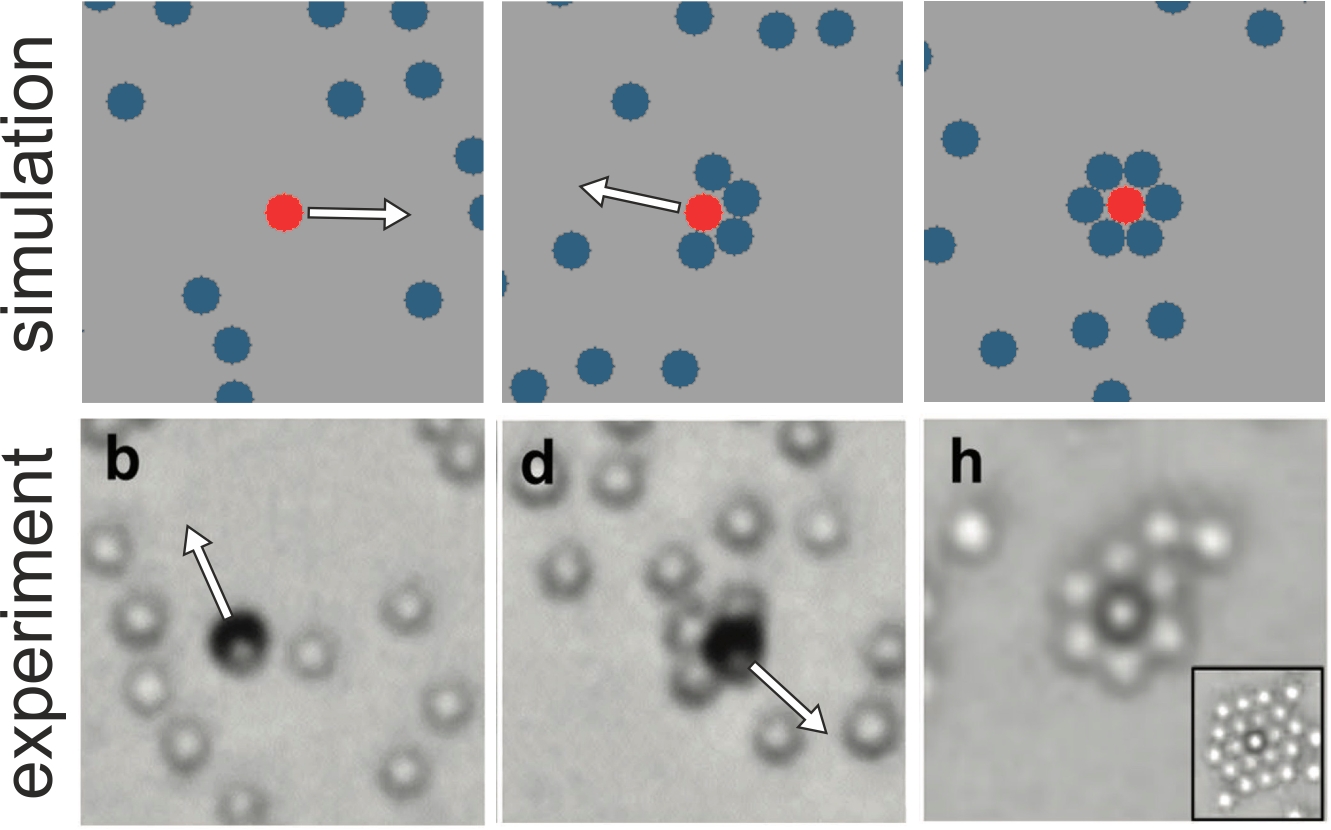}
  \caption{Velocity-reversals in simulations (top row) and experiments from ref.~\cite{Singh} (bottom row). 
  The active particles (red dots in the simulation; dark particles in the experiment) are initially propelling 
  with their catalytic cap ahead. Both, simulation and experiment show a preferred aggregation 
  of passive particles at the active particles' catalytic caps, followed by a velocity reversal. 
  Parameters have been matched to experiments (see Appendix) leading to the following values: 
  $K_d\sim 5\cdot 10^3$, $K_0=250$, $\mathcal{B}_p=43654$, $\mathcal{B}_a=1284$, ${\tilde{d}}=1.0$, $\mathcal{D}_r=1.0$, $\mathcal{D}_t=\frac{1}{3}$, $\mathcal{D}_c=4444.0$ and $\phi_a = 0.4\%$, $\phi_p = 8\%$ the area fraction of active and passive particles.}
\label{PropReversion}
\end{figure}

\begin{figure}
  \centering
  \includegraphics[width=0.47\textwidth]{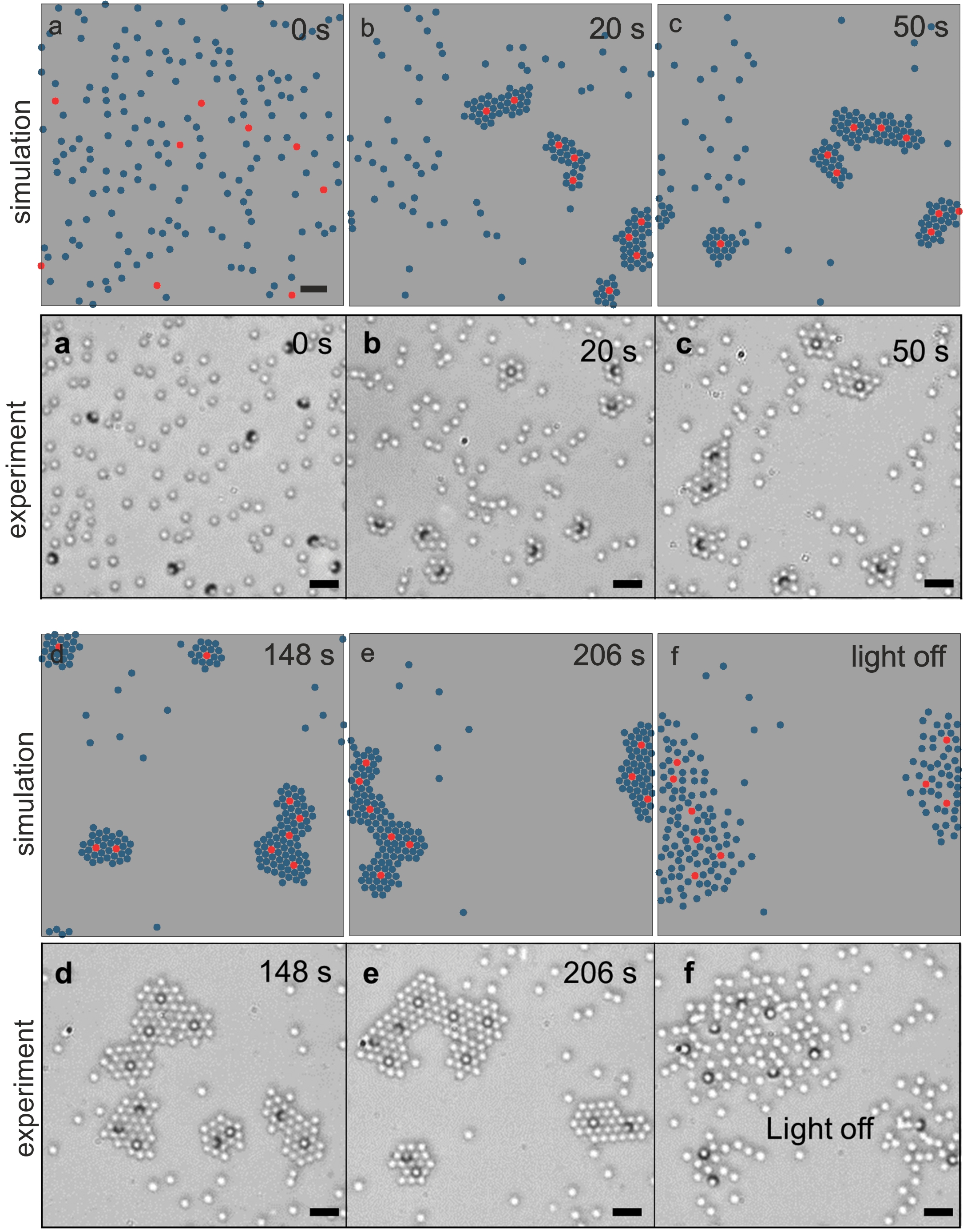}
  \caption{Simulation snapshots of passive (blue) and active particles (red) compared to experiments \cite{Singh} shown in the $x-y-$plane.  
  Parameters as in Fig.~\ref{PropReversion}.}
\label{CompSimuExo}
\end{figure}

To test our model, we first extract the relevant parameters from experiments 
(see Appendix) and perform Brownian dynamics simulations, starting from a uniform and disordered particle distribution, 
coupled to a finite difference scheme for the dynamics of the chemical field (initially uniform), using 
a 2D simulation box of size $\mathcal L = \frac{L}{d}$ with periodic boundary conditions \footnote{
We alternatingly iterate the particle positions and the chemical field by one step, using 
time step size ranges from $\Delta t=2.5\cdot10^{-7}$ to $\Delta t=2\cdot10^{-6}$
depending on the specific simulation. For the finite differences grid we chose a spacing of $\Delta x=\frac{2}{9}d$.}.
As in experiments \cite{Singh}, in the simulations, at early times, the active Janus colloids self-propel with their catalytic cap ahead, whereas the passive colloids diffuse rather randomly. 
Within a second or less, however, passive colloids move permanently towards the caps of the 
active colloids and attach there. Also as in experiments, we never observed cases of immediate active-active 
binding at the very low active particle density ($\phi_a=0.4\%$) considered.
Remarkably, once a few passive colloids have attached to an active one, the resulting aggregate reverts its direction of motion (see Figs.~\ref{Cartoon},\ref{PropReversion} and movie S1). 
In our model, the underlying mechanism is that even at contact, passive colloids, sitting at the cap of an active colloid, 
move up the gradient of the chemical, produced at the cap. Thus they push the active colloid against its self-propulsion direction, which is particularly efficient if 
several passive colloids attach to the cap of the active one and collectively push it backwards (see Fig.~\ref{PropReversion}). Notice that this phenomenon reflects the nonreciprocal character 
of phoretic interactions in nonidentical particles \cite{Ivlev2015}. 
As time proceeds the passive colloids continue aggregating around the active ones and after a few seconds only, they form complete shells 
surrounding the active core particle. 
The resulting clusters move very slowly at this stage, but continue growing by attracting further passive colloids. 
Fig.~\ref{CompSimuExo} illustrates the subsequent steps of cluster formation in comparison with experiments and shows that the clusters grow and 
merge on similar timescales as in experiments. 
Notice here, the important role of the effective screening of the phoretic interactions probably occurring due to bulk reactions, which has so-far often been neglected.
In fact, if the effective screening was absent, the clusters would move much too fast as shown in the Appendix in Fig.~\ref{FittedData}c. Here already at $t=24s$ all particles have collapsed into 
one macroscopic cluster, opposing experimental results, which show free particles even for $t>200s$. In addition, fitting the tracer motion to experimental 
measurements strongly suggests the existence of such an effective screening 
as we discuss in detail in the Appendix.

\begin{figure}
  \centering
   \subfigure[square - simulation]{\label{square}\includegraphics[width=0.21\textwidth]{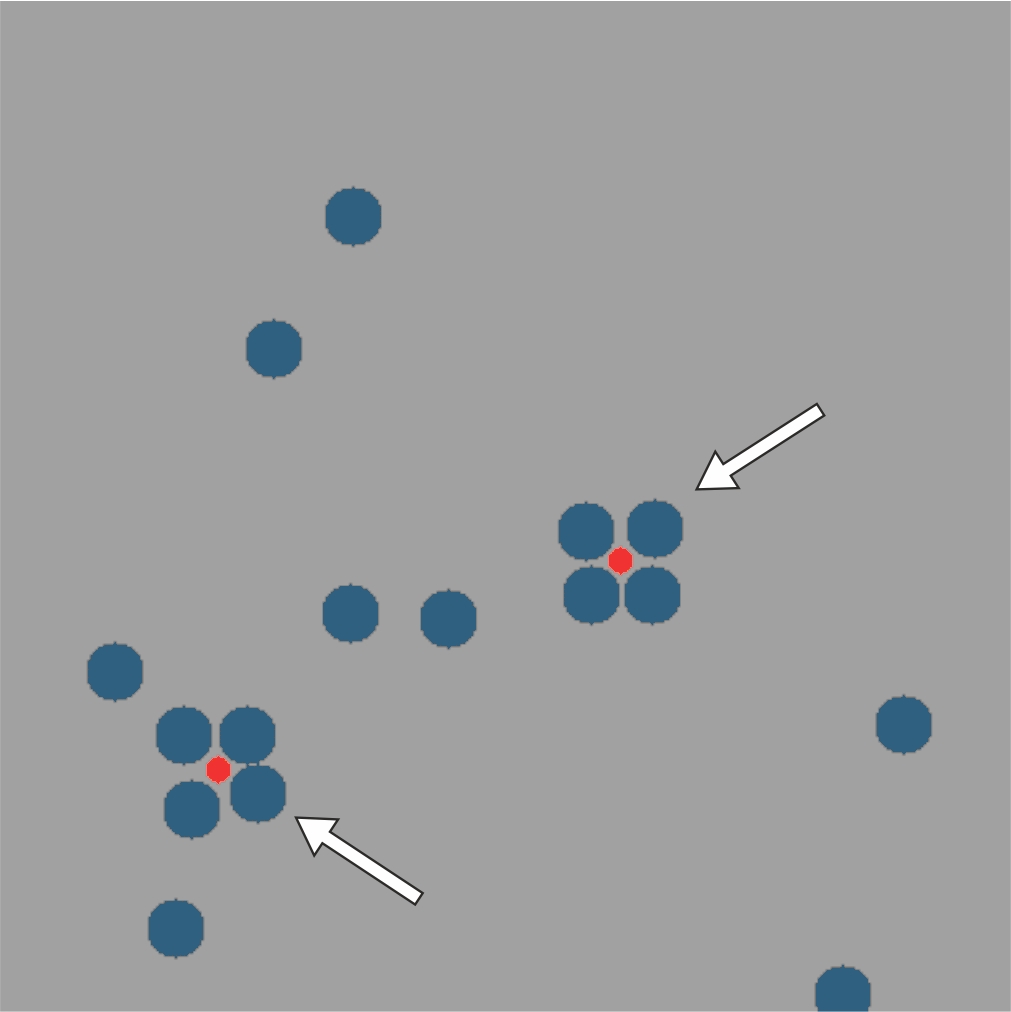}}
   \subfigure[heptagonal - simulation]{\label{hepta}\includegraphics[width=0.21\textwidth]{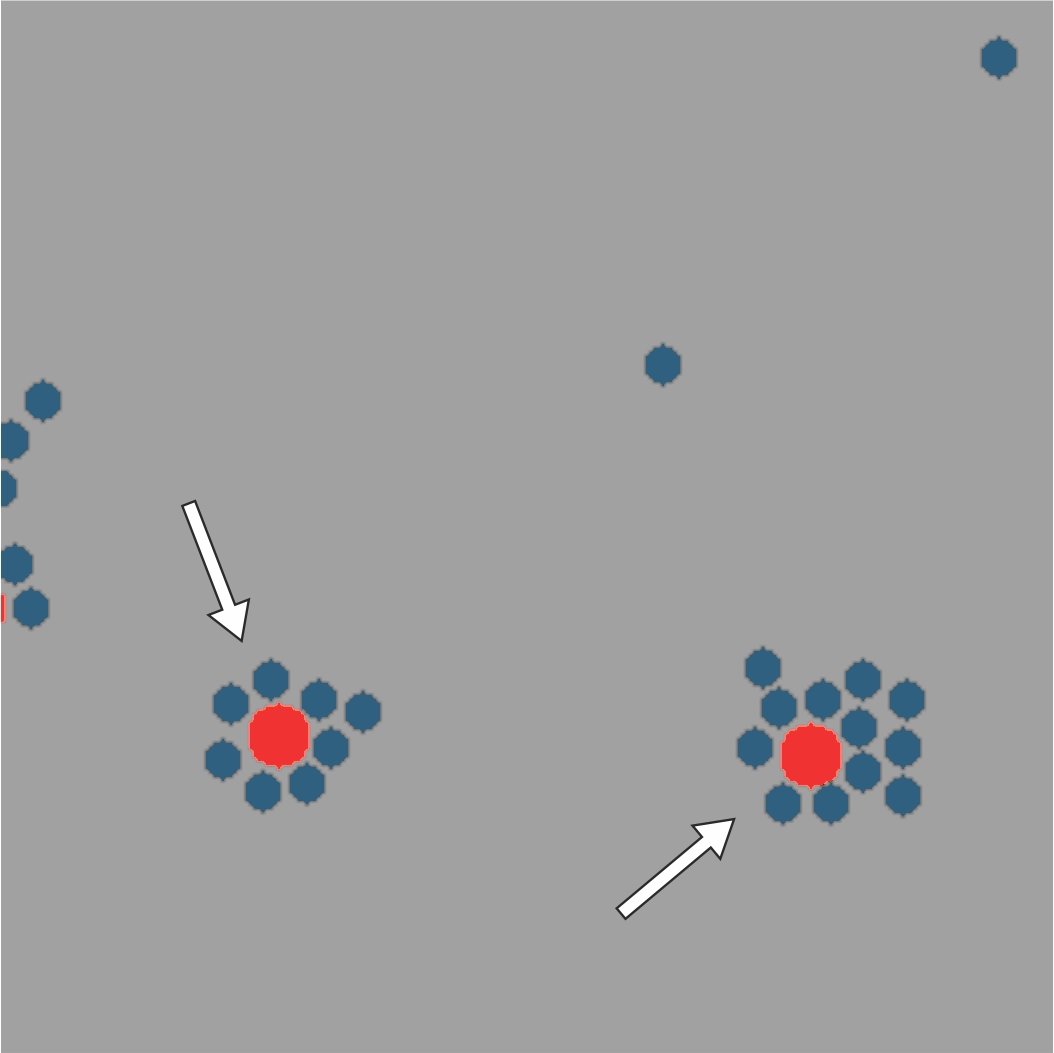}}  \\
   \subfigure[square - experiment]{\label{quadExp}\includegraphics[width=0.21\textwidth]{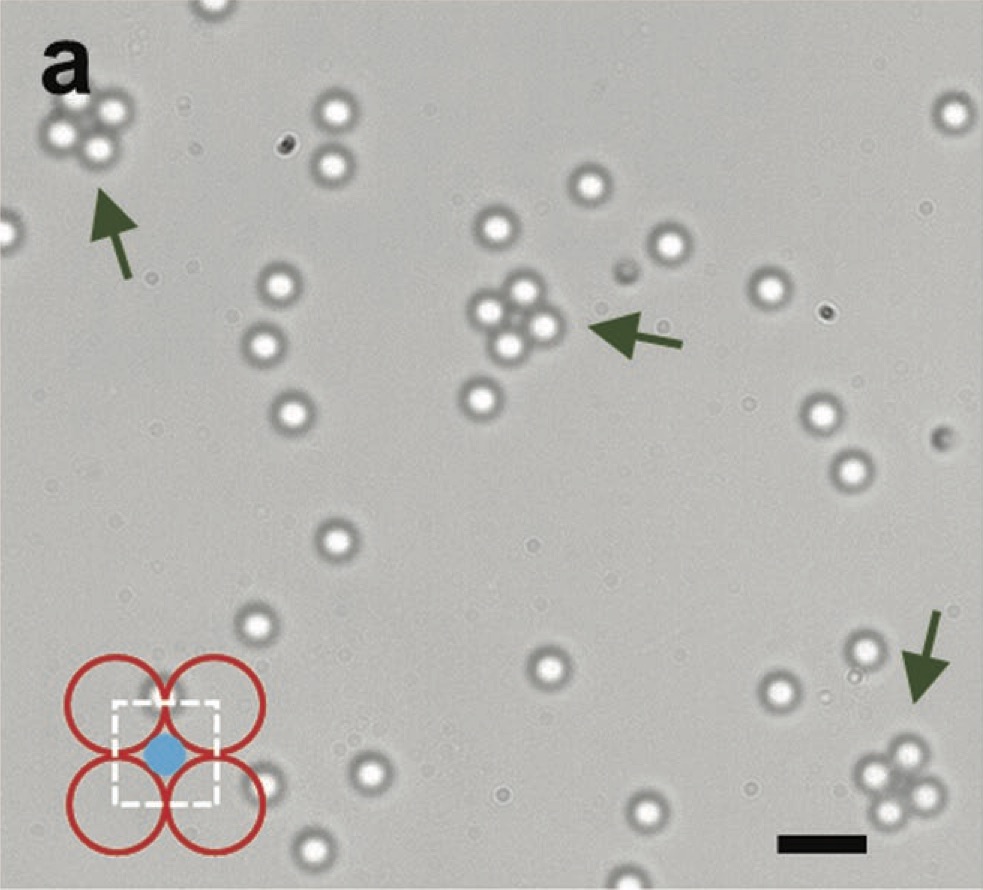}}
   \subfigure[heptagonal - experiment]{\label{heptExp}\includegraphics[width=0.21\textwidth]{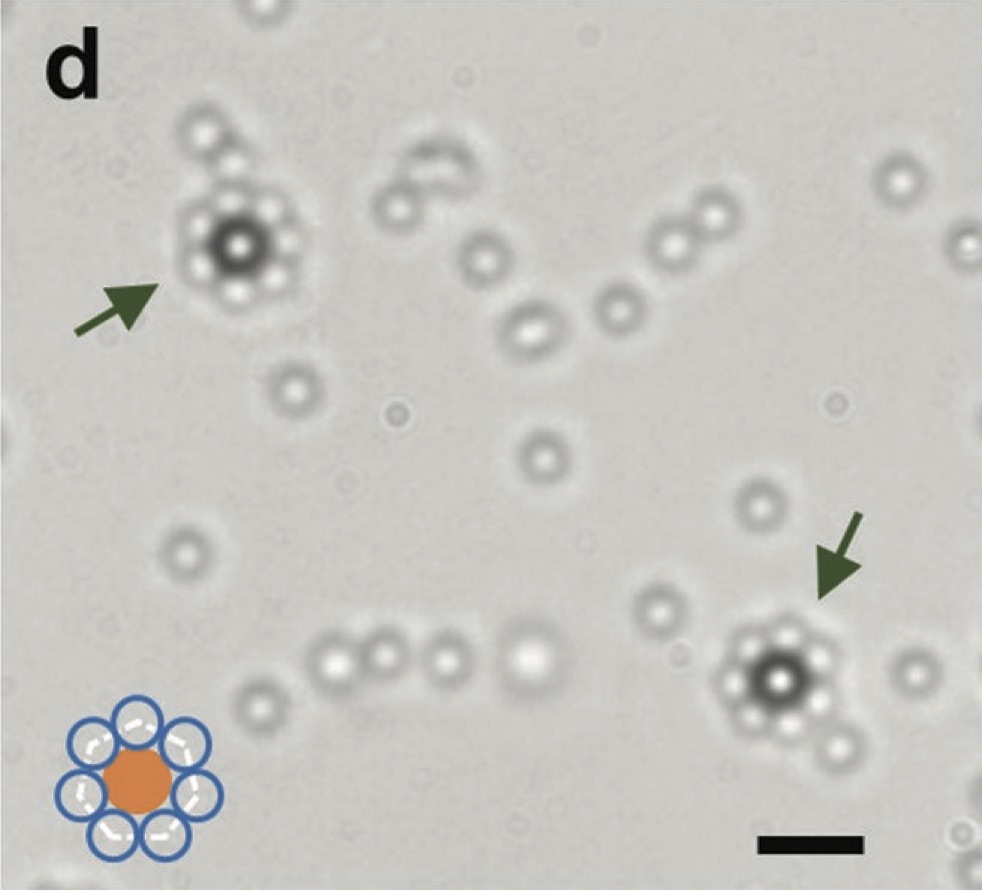}} 
  \caption{Comparison of simulations for active-passive mixtures with different particle sizes with experiments \cite{Singh}, shown in the $x-y$-plane. 
  Parameters as in Fig.~\ref{PropReversion} except for the varied particle diameters $d'$, which we have chosen as in experiments, i.e. in panels (a) and (c) for the active particles we have $d'=\frac{2}{3}d$ and $d'=1.4d$ for the passive ones ($\phi_a=0.13\%$ and $\phi_p=5.2\%$); in (b) and (d) we have $d'=d$ and  $d'=\frac{2}{3}d$ for the active and the passive particles respectively ($\phi_a=0.3\%$ and $\phi_p=1.2\%$) and for the dimensionless production rate for which we have chosen $K_0=50$, because corresponding experiments use a reduced laser power.}
\label{lattice}
\end{figure}

Within each of the clusters, the particles closely pack and show a hexagonal structure. In experiments, this changes when using active and passive particles of different size: the same is seen in our model, where we observe square-structures and heptagonal ones (see Fig.~\ref{lattice}).

\section{Chemical Repulsion}\label{chemicalrepulsion}
While recent experiments have started exploring also active-passive mixtures with repulsive phoretic interactions \cite{Wang2018,Wang2019,Chattopadhyay,WangExclusion}
this case has received somewhat less attention in the literature modeling active-passive mixtures so far. This case, where the (active) colloids phoretically move down gradients of the effective chemical field $c({\bf r},t)$ may occur e.g.~for colloids moving up the fuel-species or down the resulting chemical production. 

\begin{figure*}
  \centering
  \includegraphics[width=0.98\textwidth]{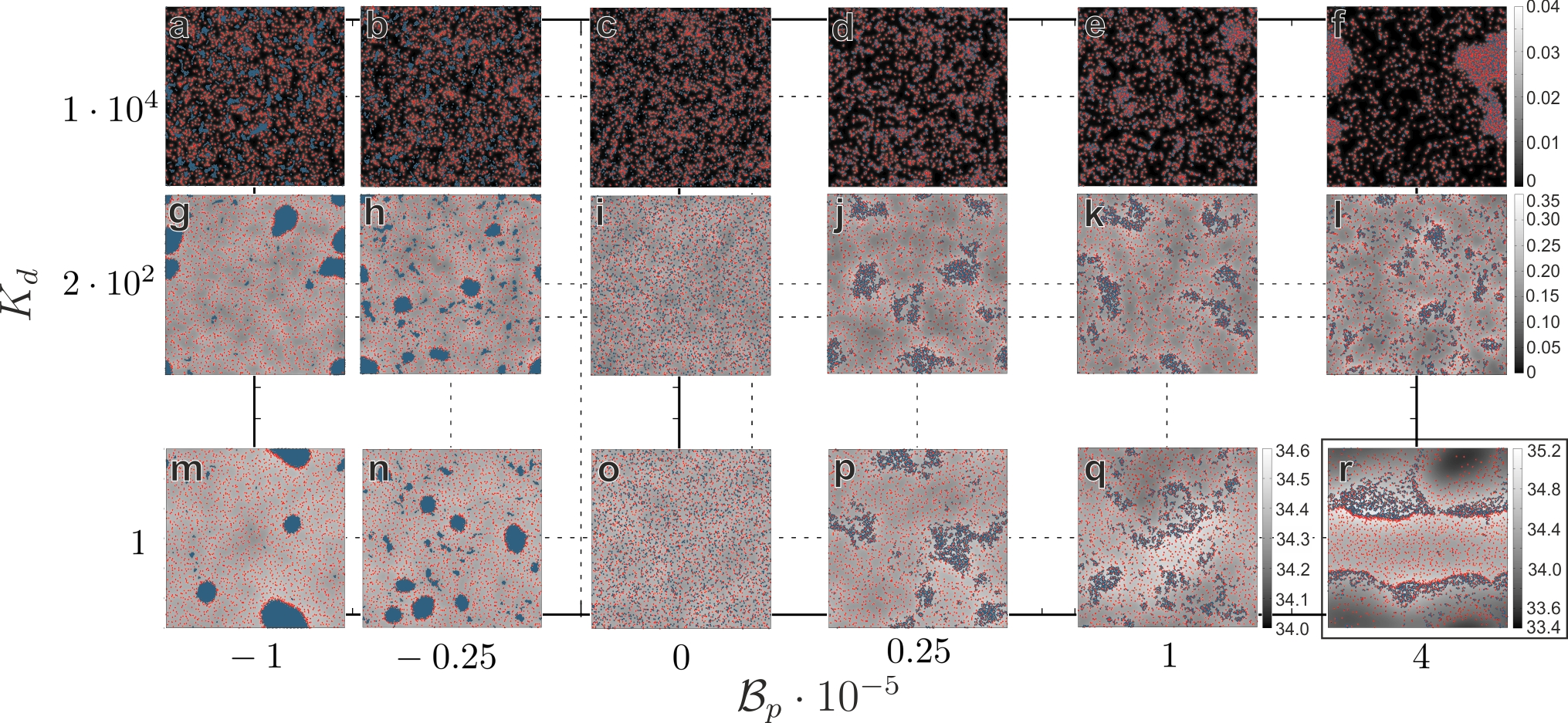}
  \caption{State diagram of a 1:1 mixture of active (red) and passive (blue) particles, based on simulation snapshots shown in the $x-y$-plane recorded at $t=18$. Parameters: $\mathcal{B}_a = -1284.0$, $N=4000$, $\mathcal{L}=120$, other parameters as in Fig.~\ref{PropReversion}. Background colors represent the chemical density in reduced units, with greyscale-bars applying to all panels left of it. 
  Note here that small $K_d$ values lead to a large chemical background density not affecting the gradients. (See also movies S3, S4, S5 and S6.)} 
\label{PhaseDiag}
\end{figure*}

We therefore switch the sign of $\mathcal{B}_a$, but keep its strength which is dictated by the self-propulsion velocity. Now 
systematically varying 
$\mathcal{B}_p$ from negative to positive values in our simulations, as well as $K_d$, which determines the range of the phoretic interactions, 
we obtain the state diagram shown in 
Fig.~\ref{PhaseDiag}.

When $\mathcal{B}_p<0$ and $K_d$ is large ($K_d\sim 10^4$), 
we observe that the passive particles aggregate in the dynamically evolving chemical density minima
(Fig.~\ref{PhaseDiag}a,b and movie S3) which resembles the structures seen \cite{Chattopadhyay} (Fig.\ 5 and 6 therein) for a mixture of 
$\chem{Cu}-\chem{TiO_2}$ and $\chem{SiO}_2$ particles.
Remarkably, when $K_d$ is significantly smaller ($K_d\sim 1$) so that the phoretic repulsions have a longer range, 
the passive particles form comparatively large and persistent clusters, 
surrounded by a dynamic shell of active particles (Fig.~\ref{PhaseDiag}g,h,m,n and movie S4).   
To understand why these clusters emerge, note that the passive particles do not produce any chemical, leading to a self-stabilization of large passive clusters: 
these clusters serve as an 'exclusion zone' for active particles and hence for chemical sources resulting in a persistent chemical density minimum 
in the cluster center attracting the chemorepulsive passive colloids. The active colloids are also attracted by the 
chemical density minimum, but weaker than the passive ones and therefore form a diffusive corona around the passive aggregate. 

If $\mathcal{B}_p>0$ so that the active particles effectively attract the passive ones but mutually repel each other, 
we observe a completely different phenomenology. Here, active and passive particles have a tendency to 'mix', which is particularly 
pronounced if $\mathcal{B}_p$ and $K_d$ are both large so that the interactions are rather strong and short ranged (Fig.~\ref{PhaseDiag}e,f and Movie S5).
Such clusters occur because the passive particles aggregate around the active ones and are attracted over rather long distances, which overcompensates the 
active-active repulsion leading to an overall aggregation. 

Interestingly, when lowering $K_d$ so that the phoretic repulsions are longer ranged, we observe dynamic patterns involving clusters which dynamically form and fragmentate and sometimes also fronts 
made of a mixture of active and passive particles moving through an almost uniform gas of active particles (Fig.~\ref{PhaseDiag}r and movie S6). 
Why do these patterns occur rather than stationary clusters as for large $K_d$?
For small $K_d$ in aggregates of active and passive particles
the active particles repel each other over comparatively long distances, opposing aggregation at some point. 
The crucial point is now, that there is generally a delay between the chemical production and the response 
of the colloids, so that the aggregation may overshoot, before the chemical density, which is responsible for the mutual repulsions, follows, also overshoots and repels particles, fragmentating the cluster. 
(This mechanism is similar to the delay-induced instability which has been derived in the framework of a linear stability
analysis for active colloids in refs.~\cite{Liebchen2015,Liebchen2017} and has been shown to initiate traveling waves.)

\section{Conclusions}\label{conclusion}
The explored model for active-passive colloidal mixtures with phoretic attractions and parameters being matched to experiments \cite{Singh} describes the aggregation of passive colloids around active 
seeds, including the characteristic 
velocity reversals of the clusters occurring in the coarse of their formation. In our model the dynamical velocity-reversals emerge due to unidirectional
phoretic attractions of the passive colloids towards the caps of the active colloids, which take place even at close contact and allow the passive
particles to push the active ones back, against their natural self-propulsion direction. That is, the velocity-reversals seen in experiments \cite{Singh} hinge on the anisotropy and nonreciprocity of phoretic interactions. 

Our results suggest that the aggregation dynamics is only described correctly when accounting for chemical decay processes leading to an 
effective screening of phoretic interactions
at a scale of a few particle sizes (due to bulk reactions). This is in agreement with measurements of the phoretic attractions, also suggesting the presence of an effective screening. 

Finally, for mixtures involving repulsive phoretic interactions, which have received less attention in the literature so-far, 
we predict a range of patterns, such as dynamical aggregation of passive colloids, core-shell clusters and moving patterns. 
These results could inspire new experiments to explore parameter regimes involving repulsive phoretic interactions. 

Overall, the present work shines light on the importance of the unusual properties of phoretic interactions, such as their anisotropy and nonreciprocity as well on their effective screening, in active-passive colloidal mixtures. This might be useful information for future more precise models accounting for various aspects which we have neglected in the simple, effective model considered in the present work. This may include systematic description of phoretic near-field interactions, and accounting for hydrodynamic interactions 
and effects due to boundaries (substrates).  


%


\section{Acknowledgements}
We thank Juliane Simmchen and LinLin Wang for helpful discussions and Dhruv Singh for providing the experimental data. Financial support within SPP 1726 of DFG is acknowledged.

\section{Appendix\label{appA}}
To be realistic, we now estimate most of the parameters based on the experiment \cite{Singh}.
Since self-propulsion and chemical interactions among the active colloids are inseparably linked with each other \cite{Liebchen2017}, 
we can fix $\beta_a$ by matching the self-propulsion velocity of a single swimmer with experiments:  
Using $v_0=6.0 \,\mu m/s$ \cite{Singh} (at ``full'' laser intensity $P_\text{max} = 320 \,\text{mW}/\text{cm}^2$), 
we obtain $\beta_a \approx 1300 \,\mu \text{m}^4/\text{s}$. The product $\beta_p\cdot k_0$
follows from measurements of the speed of passive colloidal tracers towards an active Janus colloid in \cite{Singh}. 
By fitting the experimental data (see below) we obtain 
$\beta_p\cdot k_0 \approx 2\cdot 10^6 \,\mu\text{m}^{D+2}/\text{s}^2$ (and choose $\beta_p=44.200 \mu \text{m}^4/s$ and $k_0=50\,\frac{1}{\text{s}}$) and also $k_d\sim 10^3/s$.
We finally estimate the 
chemical diffusion as $D_c \sim 2\cdot 10^3 \,\frac{\mu\text{m}^2}{\text{s}}$. 
Using the particle diameter $d=1.5\,\mu$m and the rotational diffusion coefficient $D_r=0.2\,\frac{1}{\text{s}}$, both as in \cite{Singh}, these parameter values
result in the following dimensionless numbers: 
$K_d\sim 5\cdot 10^3$, $K_0=250$, $\mathcal{B}_p=43654$,  
$\mathcal{B}_a=1284$ and $\mathcal{D}_c=4444.0$.
Finally, following the Stokes-Einstein relations for translational and rotational diffusion we obtain $\mathcal{D}_t=1/3$. \\

\begin{figure}
  \centering
  \includegraphics[width=0.47\textwidth]{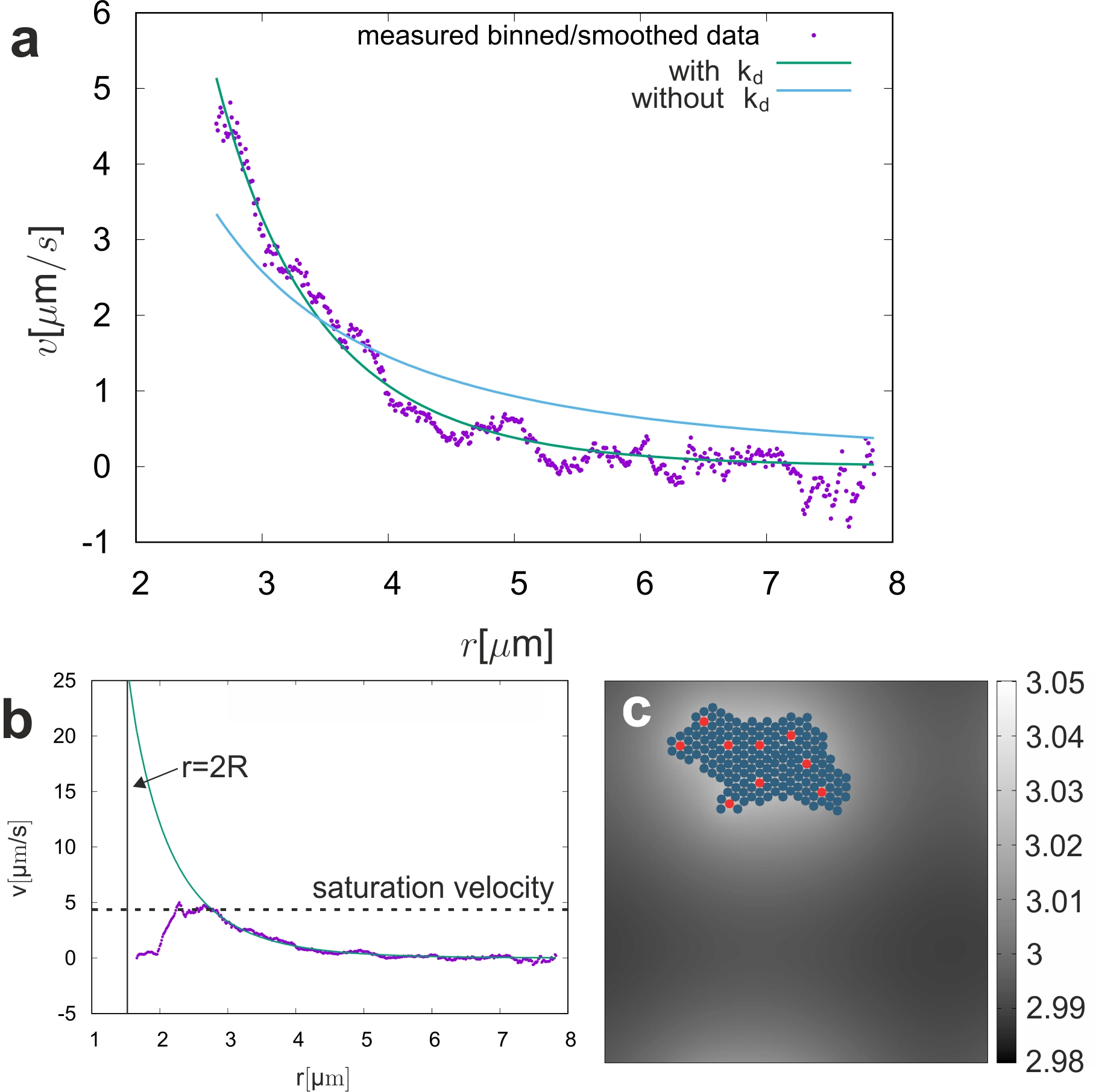}
  \caption{a) 
  Radial velocity of passive particles as a function of the center-to-center distance to an attractive active particle. 
  The purple dots represent averaged, binned and smoothed experimental data. 
  Blue and green lines represent nonlinear least-square Marquardt-Levenberg
  fits of (Eq.~\ref{vr}) for $k_d=0$ (unscreened) and for $k_d \neq 0$ (effectively screened) respectively. 
  b) Full data set including near field. 
  c) Exemplaric simulation snapshot for the same parameters underlying Fig.~\ref{PropReversion} but with $k_d=0$ 
  leading to a rapid collapse of the system into a single cluster already at $t=24s$. This cluster moves significantly faster than 
  the clusters in Fig.~\ref{CompSimuExo} because of the asymmetric distribution of active particles.
  Note here that the chemical background concentration is much higher than in Fig.~\ref{CompSimuExo}, because $k_d=0$, but 
  the gradients are comparable.}
\label{FittedData}
\end{figure}

Let us now discuss how in detail we estimate $\beta_p$ and $k_d$ from experiments \cite{Singh} where the motion of passive tracers in the field produced by 
an active particle has been measured (see also \cite{Liebchen2019} for a preliminary discussion). In this section we use dimensional 
parameters to simplify the comparison to experiments. 
The steady state for the 3D chemical field due to a single point source at the origin is described by 
$0=D_c \nabla^2 c + k_0 \delta({\bf r})-k_d c$ which has the following solution \cite{BookChapter}
\begin{equation} \label{eq:II1}
c(r) = \frac{k_0}{4\pi D_c r}{\rm e}^{-\kappa r} \quad \text{with} \quad \kappa = \sqrt{k_d/D_c}
\end{equation}
The speed of passive (tracer) colloids in a chemical field which varies slowly on scales comparable to the particle size is $v(r)\approx \beta_p \nabla c$ 
yielding: 
\1  
v(r)=-\frac{\beta_{\text{p}} k_0 {\rm e}^{-\kappa r}}{4 \pi D_c r} \left(\frac{1}{r}+\kappa\right)
\label{vr}
\2
In Fig.~\ref{FittedData} we have fitted this expression, both for $\kappa=0$ and for $\kappa\neq 0$ to the same experimental raw data which 
underlies Fig.\ 3 in \cite{Singh}, 
for the radial velocities of pairs of one active and passive particle (the active particle being ``immobilized'' via the substrate)
as a function of the 
center-to-center distance $r \in (2.5-8\,\mu \text{m})$. 
We have slightly averaged and smoothed the corresponding data. 

Performing a nonlinear least-square Marquardt-Levenberg algorithm fit we obtain 
$B=\beta_p\cdot k_0=2.21\cdot 10^6 \,\mu\text{m}^{D+2}/\text{s}^2$ and $k_d = 1148 \,\frac{1}{\text{s}}$,
as the best possible fit (green line) to the experimental data (purple dots). 
For comparison, we also show the best possible fit for $\kappa=0$, which leads to the blue line in Fig. \ref{FittedData}. 
This indicates that phoretic interactions are effectively screened in the experiment \cite{Singh}
leading to an effective range of the relevant phoretic interactions of one or a few particle diameters. 
In addition, as discussed in the main text, when assuming unscreened phoretic interactions instead, our simulations do not sensibly reproduce experimental results but lead
to a much too fast cluster growth (Fig. \ref{FittedData}c). 

Fig.~\ref{FittedData}b shows $v(r)$ including the near-field regime, where the relative velocity between the colloids saturates and decays to almost zero 
for very small distances. 
To account for this rather involved near-field behaviour in a minimal way, we set the magnitude of the chemical gradient seen by a passive colloid to a value leading to the saturation speed of $v=4.5\mu$m/s as indicated in Fig.~\ref{FittedData}b, whenever it is closer than a distance of $2R$ to an active colloid, but keep the direction of the gradient.

\subsection{Movies}
Movies S1 and S2 correspond to Figs.~\ref{CompSimuExo} Fig.~\ref{FittedData}(c) respectively and 
Movies S3, S4, S5, S6 correspond to Fig.~\ref{PhaseDiag}b,n,f,r. Parameters are the same as for the corresponding figures.


%
%

%


\bibliography{bibfilePaper}

\end{document}